%% file: main.tex
\documentclass[copyright,creativecommons]{eptcs}
\providecommand{\event}{SYNT 2016} 
\usepackage{amsthm}
\usepackage{amssymb}
\usepackage{mathtools}
\usepackage{temporal}
\usepackage{enumerate} 
\usepackage{color}
\usepackage{tikz}
\usepackage{paralist}
\usepackage{fixme,manfnt,mparhack}
\usepackage[T1]{fontenc}
\usepackage{enumerate}
\usepackage{xspace}
\usepackage{amssymb}
\usepackage{paralist}
\usepackage{wrapfig}
\usepackage{bbm}
\usepackage{cite}
\usepackage{booktabs}
\usepackage{array}
\usepackage{arydshln}

\usepackage{macros}

\providecommand{\thisvolume}[1]{this volume of {\sl Electronic
  Proceedings in Theoretical Computer Science}}

\definecolor{mygreen}{rgb}{0,0.6,0} 
\definecolor{mygray}{rgb}{0.9,0.9,0.9} 
\definecolor{mymauve}{rgb}{0.58,0,0.82}

\usepackage{listings}
\usepackage[margins]{trackchanges}
\addeditor{SJ}

\usepackage{xspace}

\input{preamble}

\title{The 3rd Reactive Synthesis Competition (\syntcomp 2016): Benchmarks, Participants \& Results}
\author{Swen Jacobs
\institute{Saarland University\\Saarbr\"ucken, Germany}
\and
Roderick Bloem
\institute{Graz University of Technology \\ Graz, Austria}
\and 
Romain Brenguier
\institute{University of Oxford\\ Oxford, UK}
\and 
Ayrat Khalimov
\institute{Graz University of Technology \\ Graz, Austria}
\and 
Felix Klein
\institute{Saarland University\\Saarbr\"ucken, Germany}
\and
Robert K\"onighofer
\institute{Graz University of Technology \\ Graz, Austria}
\and 
Jens Kreber
\institute{Saarland University\\Saarbr\"ucken, Germany}
\and
Alexander Legg
\institute{Data 61, CSIRO (formerly NICTA) and UNSW\\Sydney, Australia}
\and
Nina Narodytska
\institute{Samsung Research America\\Mountain View, USA}
\and
Guillermo A. P\'erez 
\institute{Universit\'e Libre de Bruxelles\\ Brussels, Belgium}
\and 
Jean-Fran\c{c}ois Raskin
\institute{Universit\'e Libre de Bruxelles\\ Brussels, Belgium}
\and
Leonid Ryzhyk
\institute{Samsung Research America\\Mountain View, USA}
\and 
Ocan Sankur 
\institute{CNRS, Irisa\\Rennes, France}
\and 
Martina Seidl
\institute{Johannes-Kepler-University\\ Linz, Austria}
\and 
Leander Tentrup
\institute{Saarland University\\Saarbr\"ucken, Germany}
\and 
Adam Walker
\institute{Independent Researcher}
}
\def\titlerunning{\syntcomp 2016}
\def\authorrunning{S. Jacobs et al.}
\begin{document}
\maketitle

\begin{abstract}
We report on the benchmarks, participants and results of the third reactive synthesis competition (\syntcomp 2016). The benchmark library of \syntcomp 2016 has been extended to benchmarks in the new LTL-based \emph{temporal logic synthesis format} (TLSF), and $2$ new sets of benchmarks for the existing AIGER-based format for safety specifications. The participants of \syntcomp 2016 can be separated according to these two classes of specifications, and we give an overview of the $6$ tools that entered the competition in the AIGER-based track, and the $3$ participants that entered the TLSF-based track. 
We briefly describe the benchmark selection, evaluation scheme and the experimental setup of \syntcomp 2016. Finally, we present and analyze the results of our experimental evaluation, including a comparison to participants of previous competitions and a legacy tool. 
\end{abstract}

\input{intro}

\input{benchmarks}

\input{execution}

\input{participants}

\input{results}

\input{results-tlsf}

\input{conclusions}

\bibliographystyle{eptcs}
\bibliography{synthesis}
\end{document}

%% file: preamble.tex
\newcommand{\myparagraph}[1]{\smallskip \noindent {\em #1.}}

\newcommand{\available}[1]{is available at \url{#1}. Accessed August 2016.}

\newcommand{\demiurge}{\textsf{Demiurge}\xspace}

\newcommand{\syntcomp}{SYNTCOMP\xspace}

\newcommand{\abssynthe}{AbsSynthe\xspace}
\newcommand{\simpleBDD}{Simple BDD Solver\xspace}
\newcommand{\realizer}{Realizer\xspace}
\newcommand{\termitesat}{TermiteSAT\xspace}
\newcommand{\acaciaforaiger}{Acacia4Aiger\xspace}
\newcommand{\bosy}{BoSy\xspace}
\newcommand{\party}{\textsc{Party}-Elli\xspace}
\newcommand{\sdf}{SDF\xspace}
\newcommand{\safetysynth}{SafetySynth\xspace}
\newcommand{\unbeast}{\textsc{Unbeast}\xspace}

\newcommand{\syfco}{SyFCo\xspace}

\newcommand{\Abc}{\textsf{ABC}\xspace}

\newcommand{\subby}{\kern-0.2em\leftarrow\kern-0.2em}

\newcommand{\grayed}[1]{\textcolor{gray}{#1}}

%% file: intro.tex
\section{Introduction}
\label{sec:intro}

Since the definition of the problem more than 50 years ago~\cite{Church62}, the automatic synthesis of reactive systems from formal specifications is one of the major challenges of computer science. Research into the basic questions related to the problem has led to a large body of theoretical results, but their impact on the practice of system design has been rather limited.
To increase the impact of theoretical advancements in synthesis, the reactive synthesis competition (\syntcomp) has been founded in 2014~\cite{SYNTCOMP14}. The competition is designed to foster research in scalable and user-friendly implementations of synthesis techniques by establishing a standard benchmark format, maintaining a challenging public benchmark library, and providing a \emph{dedicated and independent} platform for the comparison of tools under consistent experimental conditions.
	

The first \syntcomp was held in 2014~\cite{SYNTCOMP14}, and the second in 2015~\cite{SYNTCOMP15}. The competition is regularly associated with the International Conference on Computer Aided
Verification (CAV) and the Workshop on Synthesis (SYNT), and the results of the annual competitions are first presented at these venues.
A design choice for the first two competitions was to focus on safety properties
  specified as monitor circuits in an extension of the AIGER format known from the hardware model checking competition~\cite{HWMCC14,SYNTCOMP-format}. \syntcomp 2016 introduces the first major extension of the competition: in addition to the existing competition track, we introduce a new track that is based on properties in full linear temporal logic (LTL), given in the \emph{temporal logic synthesis format} (TLSF) recently introduced by Jacobs, Klein and Schirmer~\cite{JacobsK16}.
	
The organization team of \syntcomp 2016 consisted of R. Bloem and S. Jacobs, with technical assistance from J. Kreber for the setup and execution, and from F. Klein for the integration of TLSF.

The rest of this paper describes the design, benchmarks, participants, and
results of \syntcomp 2016. We present the benchmark set for \syntcomp 2016 in Section~\ref{sec:benchmarks}, followed by a description of the setup, rules and execution of the competition in Section~\ref{sec:setup}. In Section~\ref{sec:participants} we give an overview of the participants of \syntcomp 2016, focusing on changes compared to last year's participants. Finally, the experimental results are presented and analyzed in Section~\ref{sec:results}.

Note that more details on the goals and design of the competition can be found in the sister paper that discusses the design of \syntcomp in $2016$ and the future~\cite{JacobsB16}.

%% file: benchmarks.tex
\section{Benchmarks}
\label{sec:benchmarks}

In this section, we describe the benchmark library for \syntcomp 2016. We start by describing benchmarks in the new TLSF format, for the LTL synthesis track. Then, we describe new benchmarks in the AIGER format, followed by a summary of the classes of benchmarks in AIGER format that have already been used in \syntcomp 2014 and 2015.

\subsection{TLSF Benchmarks}
\label{sec:benchmarks-tlsf}

All of the following benchmarks have been translated into TLSF or directly encoded in TLSF by F. Klein, except for the last set of benchmarks, that has been encoded by S. Jacobs.

\paragraph{Lily benchmark set.}
This set of benchmarks was originally included with the LTL synthesis tool \textsc{Lily}\footnote{\textsc{Lily} is available at \url{http://www.iaik.tugraz.at/content/research/design_verification/lily/}. Accessed August 2016.}~\cite{JobstmannB06}. It includes $24$ benchmarks. 

\paragraph{Acacia benchmark set.}
This set of benchmarks was originally included with the LTL synthesis tool Acacia+\footnote{Acacia+ is available at \url{http://lit2.ulb.ac.be/acaciaplus/}. Accessed August 2016.}~\cite{bbfjr12}. It includes $65$ benchmarks.

\paragraph{Parameterized detector.}
This benchmark specifies a simple component that gets a number of inputs and raises its single output infinitely often if and only if all of the inputs are raised infinitely often. Parameterized in the number of inputs.

\paragraph{Parameterized arbiters.}
This set of benchmarks includes $4$ different arbiter specifications, where a central controller receives requests and gives (mutually exclusive) grants to a number of communication partners, called \emph{masters}. The basic specification is the \emph{simple arbiter}, which only requires mutual exclusion and request-response. Additionally, there are $3$ advanced specifications that require additional properties: \emph{prioritized arbiter}, \emph{round-robin arbiter}, and \emph{full arbiter} (including absence of spurious grants). All benchmarks are parameterized in the number of masters $n$ that the arbiter needs to serve.

\paragraph{Parameterized AMBA bus controller.}
The AMBA bus controller is essentially a very advanced arbiter that includes additional features, like locking the bus for a fixed or arbitrary number of steps. It has been translated from an industrial specification in natural language\footnote{AMBA Specification Rev 2.0, available at \url{http://www.arm.com/}. Accessed August 2016.} to LTL by Jobstmann~\cite{Jobstmann07b}, and is one of the most challenging and widely used benchmark for synthesis tools~\cite{BloemJPPS12,GodhalCH13,FinkbeinerJ12,BloemJK14a}. Like the arbiters, the benchmark is parameterized in the number of masters that the controller has to serve.

\paragraph{Parameterized load balancer.}
This benchmark considers a load balancer that receives jobs and distributes them to a fixed number of servers, with a number of additional properties like prioritization of the first server, and that the load is balanced between servers. The original specification has been designed by R\"udiger Ehlers~\cite{Ehlers12} for the evaluation of synthesis tool \unbeast~\cite{Ehlers11}. The benchmark is parameterized in the number of servers that can handle the jobs.

\paragraph{Parameterized generalized buffer.}
This benchmark specifies a family of buffers that transmit data from a number of senders to two receivers, based on a handshake protocol between the buffer and the other components, and an interface to a FIFO queue that is used to store data. The benchmark is parameterized in the number of senders. Like the AMBA case study, it has first been translated to LTL by Jobstmann~\cite{Jobstmann07b}.  

\paragraph{Parameterized LTL to B\"uchi translation (LTL2DBA).}
This set of benchmarks uses the synthesis procedure to generate deterministic B\"uchi automata that correspond to an LTL specification. It is based on a number of parameterized LTL formulas, as described by Tian et al.~\cite{TianSDD15}.

\subsection{AIGER Benchmarks}
\label{sec:benchmarks-aiger}

\paragraph{HWMCC benchmarks.}
This class of benchmarks is based on a subset of the benchmarks from HWMCC 2012~\cite{HWMCC14}. The idea is to consider verification benchmarks consisting of a system and a safety specification that is not satisfied by the system, and ask the question whether we can synthesize a controller for a given subset of the inputs such that the specification is satisfied. 

Benchmarks in this set are based on the unsafe instances from the \emph{single safety property} track of HWMCC 2012, modified such that between $1$ and $8$ inputs of the circuit are defined to be controllable. The files retain their original filenames, with \texttt{\_c0toc}x appended to the filename if inputs $0$ to $x$ are declared as controllable. E.g., the benchmark based on \texttt{6s5.aag}, where $8$ inputs have been declared as controllable, is named {6s5\_c0to7.aag}.

This benchmark set contains $280$ benchmarks. The original verification benchmarks have been taken from the HWMCC website\footnote{HWMCC website: \url{http://fmv.jku.at/hwmcc/}. Accessed August 2016.} and modified by S. Jacobs.

\paragraph{(Bounded) LTL to B\"uchi and LTL to parity translation (LTL2DBA/LTL2DPA).}
This class of benchmarks is based on the benchmarks for LTL to B\"uchi translation mentioned in Section~\ref{sec:benchmarks-tlsf}, but for some instances also considers the synthesis of parity automata. In addition to the original parameterization, they are parameterized in the liveness-to-safety approximation. This set contains $62$ instances and have been translated from TLSF to AIGER by G. A. P\'erez.

\paragraph{Existing benchmark library.}
The existing library that was the basis for \syntcomp 2014 and 2015 contains $3105$ benchmark instances. The following list of benchmarks was collected for the first competition, and more details can be found in the \syntcomp 2014 report~\cite{SYNTCOMP14}:
\begin{itemize}
\item Toy Examples: this benchmark set contains specifications for a number of basic building blocks of circuits, such as an adder (\texttt{add}), a bitshifter (\texttt{bs}), a counter (\texttt{count}), or a multiplier (\texttt{mult}). All of the benchmarks are parameterized in the bitwidth of the input, resulting in $176$ problem instances.
\item AMBA: a version of the bus controller specification for AMBA (as described by Bloem et al.~\cite{BloemJPPS12}), parameterized in three dimensions (number of masters, and the type and the precision of the liveness-to-safety approximation). The benchmark set contains $952$ instances.
\item Genbuf: a version of the generalized buffer specification (as described by Bloem et al.~\cite{BloemJPPS12}), parameterized in the same way as the AMBA benchmarks above. The set contains $866$ instances.
\item LTL2AIG: these are several sets of benchmarks that are based on the benchmark set of synthesis tool Acacia+~\cite{bbfjr12}, translated using the LTL2AIG tool~\cite{SYNTCOMP14}:
\begin{itemize}
\item  $50$ \texttt{demo} problems originally supplied with LTL synthesis tool \textsc{Lily},
\item $41$ \texttt{ltl2dba} and \texttt{ltl2dpa} problems that use the synthesis tool to obtain a deterministic B\"uchi automaton (dba) or a deterministic parity automaton (dpa) that corresponds to a given LTL formula, \item $42$ \texttt{gb} benchmarks, based on the generalized buffer specification, but with a notably higher difficulty than the version mentioned above,
\item $64$ load balancer benchmarks, a modification of those originally presented with LTL synthesis tool \textsc{Unbeast}\footnote{\unbeast is available at \url{http://www.react.uni-saarland.de/tools/unbeast/}. Accessed August 2016.}~\cite{Ehlers11}.
\end{itemize}
\item Factory Assembly Line: this benchmark set describes a controller for two robot arms on an assembly line. It consists of $15$ problem instances.
\item Moving Obstacle: this benchmark set describes a moving robot that should avoid a moving obstacle in two-dimensional space. It consists of $16$ problem instances.
\end{itemize}
While all classes above have been introduced for \syntcomp 2014, some of the instances have only been generated for \syntcomp 2015. Additionally, the following classes have been added for the second competition, and have been described in the \syntcomp 2015 report~\cite{SYNTCOMP15}:
\begin{itemize}
\item Washing Cycle Scheduler: specifications for a controller of a washing system, with water tanks that share pipes and cycles that can be launched in parallel. Parameterized in the number of tanks, the maximum reaction delay, and the shared water pipes. This benchmark set contains $321$ instances.
\item Driver Synthesis:
this class of benchmarks is specifies a driver for a hard disk controller with respect to a given operating system model (as described by Ryzhyk et al.~\cite{RyzhykWKLRSV14}). It is parameterized in the level of data abstraction, the precision of the liveness-to-safety approximation, and the simplification of the specification circuit by \Abc~\cite{abc}. This benchmark set contains $72$ instances.
\item Huffman Encoder:
this class of benchmarks specifies a given Huffman decoder, for which a suitable encoder should be synthesized (as described by Khalimov~\cite{Khalimov15}). The benchmark is parameterized in the liveness-to-safety approximation, resulting in $5$ problem instances.  
\item HWMCC:
this class of benchmarks is based on a subset of the benchmarks from HWMCC 2014~\cite{HWMCC14}, modified in the same way as the new HWMCC benchmarks mentioned above. This set contains $110$ instances.
\item HyperLTL:
this class of benchmarks is based on benchmark problems from HyperLTL model checking, as introduced in recent work by Finkbeiner et al.~\cite{FinkbeinerRS15}. The goal is to synthesize a witness formula (represented as a controller circuit) for a given HyperLTL property.
This benchmark set contains $21$ instances.
\item Matrix Multiplication:
these benchmarks specify matrix multiplication circuits, parameterized in the size of the input matrices. The basic benchmarks consider a single multiplication, and a variant models repeated multiplication with a subset of controllable inputs and an additional safety goal.
This benchmark set contains $273$ instances for basic case, and $81$ for the repeated case.
\end{itemize}

%% file: execution.tex
\section{Setup, Rules and Execution}
\label{sec:setup}

We give an overview of the setup, rules and execution of \syntcomp 2016. More details, as well as information about the reasoning behind different design choices, can be found in the sister paper by Jacobs and Bloem~\cite{JacobsB16}.

\subsection{General Rules}
\label{sec:rules}
Like in the previous year, there is a track that is based on safety specifications in AIGER format (in the following: AIGER/safety-track). In addition, this year for the first time there is a track based on full LTL specifications in TLSF (in the following: TLSF/LTL-track). Each track is divided into subtracks for \emph{realizability checking }and \emph{synthesis}, and into two execution modes: \emph{sequential} (using a single core of the CPU) and \emph{parallel} (using up to $4$ cores). We explain the rules of the competition, including evaluation of tools in the different tracks.

\paragraph{Submissions.}
Participants hand in their tools as source code, with installation instructions and a short description of the algorithms and optimizations used to solve the synthesis problem. Each author can submit up to three configurations of a tool per subtrack and execution mode. Tools are tested on a subset of the benchmarks before the competition, and we allow bugfixes in case of crashes or wrong results, if time permits.

\paragraph{Ranking Scheme.}
In all tracks, a correct answer within the timeout of $3600$s is rewarded with one point for the solver, and a wrong answer is punished by subtracting $4$ points.
In the realizability tracks, correctness is determined by the realizability information stored in the files, if they have been used in previous competitions, or on a majority vote of the tools that solve the benchmark, otherwise. In the synthesis tracks, if the specification is realizable, then solution has to be model checked. This differs based on the input format, as explained below.

\subsection{Specific Rules for TLSF/LTL-Track}

\paragraph{Input Format.}
In the TLSF/LTL-track, specifications are given in basic TLSF~\cite{JacobsK16}.
The organizers supply the \emph{synthesis format conversion} (SyFCo) tool\footnote{SyFCo is available at \url{https://github.com/reactive-systems/syfco}. Accessed August 2016.} that can be used to translate the specification to different existing specification formats. Specifications are interpreted according to standard LTL semantics, with respect to realizability as a Mealy machine.

\paragraph{Correctness of Solutions.}
In the synthesis subtrack, tools produce a solution in AIGER format if the specification is realizable. For syntactical correctness, the sets of inputs and outputs of the specification must be identical to the sets of inputs and outputs of the solution. Additionally, solutions are model checked against the specification with existing model checking tools. Only a solution that can be verified is counted as correct.

\paragraph{Legacy Tools.}
For comparison, we run legacy tool \unbeast, non-competitive, in the realizability subtrack. To this end, the TLSF specification is translated to the native input format of \unbeast, and a wrapper script converts outputs of the tool to the necessary format.

\subsection{Specific Rules for AIGER/safety-Track}

\paragraph{Input Format.}
In the AIGER/safety-track, specifications are given in the Extended AIGER Format for Synthesis~\cite{SYNTCOMP-format,SYNTCOMP14}, modeling a single safety property.

\paragraph{Correctness of Solutions.}
In the synthesis subtrack, tools must produce a solution in AIGER format if the specification is realizable. For syntactical correctness, this solution must include the specification circuit, and must define how those inputs that are declared as \texttt{controllable} are computed (for details, see the \syntcomp 2014 report~\cite{SYNTCOMP14}). 
Additionally, these solutions are also model checked, and only verified solutions are counted as correct.
To facilitate verification, synthesis tools can optionally output an inductive invariant that witnesses the correctness of the solution. If an invariant is supplied, we first try to determine correctness with an invariant check, and fall back to full model checking if the check is inconclusive.

\paragraph{Legacy Tools.}
For comparison, we run some of the entrants of \syntcomp 2014 and 
\syntcomp 2015 in the AIGER safety track. This allows us to highlight 
the progress of tools over the course of the last two years.

\subsection{Selection of Benchmarks}
\label{sec:selection}

\paragraph{AIGER/safety-track.} In the AIGER-based track, the selection of benchmarks is based on information about the realizability and difficulty of benchmark problems that has been obtained from the results of previous competitions. This information is stored inside the benchmark files, as described in the \syntcomp 2015 report~\cite{SYNTCOMP15}.
From each class of benchmarks, we selected a number of problems with an even distribution over difficulties (in terms of the ratio of solvers from previous competitions that were able to solve the benchmark).
For benchmarks that have been added in 2016, we estimated their difficulty and realizability based on preliminary experiments with solvers from \syntcomp 2015.

The number of selected problems from each category (cp. Section~\ref{sec:benchmarks}) is given in Table~\ref{tab:selected-benchmarks}. Compared to \syntcomp 2015, we 
\begin{itemize}
\item replaced the existing HWMCC benchmarks completely with a selection from the new HWMCC set, since the existing ones were very hard and most instances could not be solved at all,
\item replaced the existing \texttt{ltl2dba}/\texttt{ltl2dpa} (LTL2AIG) benchmarks with the new LTL2DBA/LTL2DPA benchmarks, since the existing instances were very easy and in 2015, all participants solved all instances,
\item reduced the number of benchmarks from the toy examples from $8$ to $5$ per example, and
\item changed the selection for some existing benchmark classes to make it more challenging.
\end{itemize}

\begin{table}[h]
\caption{Number of selected Benchmarks per Category, AIGER/safety-track}
\label{tab:selected-benchmarks}
\centering
\def\arraystretch{1.2}
\begin{tabular}{ll|ll}
Category & Benchmarks & Category & Benchmarks\\
\hline
AMBA & 16 & Genbuf (LTL2AIG) & 8\\
(Washing) Cycle Scheduling & 16 & Add (Toy Examples)& 5\\
Demo (LTL2AIG)& 16 & Count (Toy Examples)& 5\\
Driver Synthesis & 16 & Bitshift (Toy Examples)& 5\\
Factory Assembly Line & 15 & Mult (Toy Examples)& 5\\
Genbuf & 16 & Mv/Mvs (Toy Examples)& 5\\
HWMCC & 16 & Stay (Toy Examples)& 5\\
HyperLTL & 16 & Huffman Encoder & 5\\
Load Balancer (LTL2AIG)& 16\\
LTL2DBA/LTL2DPA & 16\\
Moving Obstacle & 16 & \\
Matrix Multiplication & 16 & {\bf Total:} & {\bf 234}\\
\end{tabular}
\end{table}
\noindent Like last year, in the synthesis subtracks we only considered benchmark instances that have been solved by at least one participant in the realizability track, resulting in a selection of $215$ of the instances above.

\paragraph{TLSF/LTL-track.} In the TLSF-based track, for realizability checking we have used all of the non-parameterized benchmarks, and have scaled up the parameter of the parameterized benchmarks until none of the tools was able to solve them (or we deemed it very likely that the next value could be solved by any of the tools).\footnote{Note that this makes comparison to tools that did not participate in the competition slightly non-trivial: if the additional tool can solve one of the parameterized benchmarks for higher parameters, then an approximately correct solution is to count these additional solutions towards the additional tool, while assuming that none of the other tools can solve them.} Overall, $195$ instances were used for realizability checking. For synthesis, we excluded all the benchmarks for which none of the competitors could determine realizability, resulting in an overall set of $185$ instances.

\subsection{Execution}
\label{sec:execution}
Like in the previous year, \syntcomp 2016 was run at Saarland University, on a set of identical machines with a single quad-core intel Xeon processor (E3-1271 v3, 3.6GHz) and 32 GB RAM (PC1600, ECC), running a GNU/Linux system. Each node has a local 480 GB SSD that can be used as temporary storage.

Also like in previous years, the competition was organized on the EDACC 
platform~\cite{BalintDGGKR11}, with a very similar setup.
To ensure a high comparability and reproducability of our results, a complete machine
was reserved for each job, i.e., one synthesis tool (configuration) running 
one benchmark. Olivier Roussel's 
\texttt{runsolver}~\cite{Roussel11}
was used to run each job and to measure CPU time and wall time, as well as 
enforcing timeouts. As all nodes are 
identical and no other tasks were run in parallel, no other limits than a 
timeout of $3600$ seconds (CPU time in sequential mode, wall time in
parallel mode) per benchmark was set.
Like last year, we used wrapper scripts to execute solvers that did not conform completely with the output format specified by the competition, e.g., to filter extra information that was displayed in addition to 
the specified output.

The model checker used for checking correctness of solutions for the AIGER/Safety track is IIMC\footnote{IIMC is available at \url{ftp://vlsi.colorado.edu/pub/iimc/iimc-2.0.tar.gz}. Accessed August 2016.} in version 2.0. This year, we also allowed the safety solvers to additionally output a winning region of the safety game, which should be an inductive invariant for the synthesized circuit. If this information was supplied, then we first used an invariant check to determine correctness of the solution, and used full model-checking as a fallback solution if the invariant check failed.

For the TLSF/LTL track, the model checker used was V3\footnote{V3 is available at \url{https://github.com/chengyinwu/V3}. Accessed August 2016.}~\cite{WuWLH14}.

%% file: participants.tex
\section{Participants}
\label{sec:participants}
Overall, nine tools were entered into \syntcomp 2016: six in the AIGER/safety-track, and three in the TLSF/LTL-track. We briefly describe the participants and give pointers to additional information.

\subsection{AIGER/safety-Track}

This track had six participants in 2016: \abssynthe, \demiurge and \simpleBDD were re-entered, and we received submissions of new tools \safetysynth, \sdf and \termitesat. Participants \abssynthe, \simpleBDD, \safetysynth and \sdf are based on the classical backward-fixpoint-based approach to solving safety games using BDDs, and mainly differ in the implemented optimizations and heuristics.  In contrast, \demiurge and \termitesat implement novel SAT-based approaches, inspired by corresponding approaches for finite-state model checking. An overview of the optimizations implemented in the different BDD-based tools is given in Table~\ref{tab:optimizations}. For detailed explanations of the different optimizations, we refer to the \syntcomp 2014 report~\cite{SYNTCOMP14}.\footnote{In particular, note that ``partitioned transition relation'' in our case only means that a separate transition function is used for every latch of the circuit, but does not include further partitioning or introduction of auxiliary variables for shared parts (cp.~\cite{BurchCL91}).}

\begin{table*}[h]
\caption{Optimizations implemented in BDD-based Tools.}
\label{tab:optimizations}
\centering
\def\arraystretch{1.1}
\begin{tabular}{r|ccccc}
  Technique                       & \abssynthe & \safetysynth & \sdf & \simpleBDD \\  \hline
automatic reordering              & x          & x         & x         & x \\
eager dereferencing of BDDs       &            & x         & x         & x \\
direct substitution               & x          & x         & x         & x \\
partitioned transition relation   & x          & x         & x         & x \\
simultaneous conjunction and abstraction & x   & x         & x         & x \\
compositional synthesis  					& x					 &           &           & \\
abstraction-refinement						& x				   &           &           & x \\
co-factor based strategy extraction & x        & x         & x         & N/A\\
forward reachability analysis     & x          &           &           & N/A \\
\Abc minimization                 &            & x         &           & N/A\\
additional optimizations          & x          &           & x         & x \\
\end{tabular}
\end{table*}

\subsubsection{Swiss \abssynthe v2.0}
The new version of \abssynthe was submitted by R. Brenguier, G. A. P\'erez, J.-F. Raskin, and O. Sankur, and competed in both the realizability and the synthesis track.

\abssynthe implements different BDD-based synthesis approaches, combining the standard fixpoint computation for safety games with some or all of the following:
\begin{itemize}
\item decomposition of the problem into independent sub-games, with different methods for merging the games after solving them (as described by Brenguier et al.~\cite{BrenguierPRS15} and in the \syntcomp 2015 report~\cite{SYNTCOMP15}),
\item abstraction-refinement algorithms that solve the game on overapproximations of the possible behaviors; this year, the abstraction algorithm is replaced with the one used also in \simpleBDD (see the \syntcomp 2014 report~\cite{SYNTCOMP14}), and 
\item portfolio approaches that run different algorithms (or algorithms with different settings) in parallel.
\end{itemize}

\noindent In sequential mode, three different algorithms were entered into the competition:
\begin{itemize}
\item configuration seq1 uses a standard BDD-based fixpoint computation with several optimizations (see Table~\ref{tab:optimizations}), but without compositionality or abstraction,
\item configuration seq2 uses the new abstraction algorithm, but no compositionality, and
\item configuration seq3 use a compositional algorithm, combined with an abstraction method that falls back to the concrete game if no solution is found when reaching a fixed threshold.
\end{itemize}

\noindent In parallel mode, also three different methods competed:
\begin{itemize}
\item configuration par1 runs the three sequential configurations in parallel, plus one additional configuration that uses abstraction with fixed threshold, but no compositionality,
\item configuration par2 runs four copies of seq1, only modified in the BDD reordering technique that is used (SIFT, WINDOW2, WINDOW3, or WINDOW4), and
\item configuration par3 runs four copies of seq2, with the same set of different reordering techniques.
\end{itemize}

Strategy extraction in \abssynthe uses the co-factor-based approach of Bloem et al.~\cite{BloemGJPPW07}, with some additional optimizations as described in the \syntcomp 2014 report~\cite{SYNTCOMP14}. To facilitate verification of solutions, \abssynthe also produces the winning region and includes it in the solution file.
The algorithms used in \abssynthe have been described in more detail by Brenguier et al.~\cite{BrenguierPRS14,BrenguierPRS15}.

\paragraph{Implementation, Availability.}
\abssynthe is implemented in C++, and depends on the AIGER toolset\footnote{The AIGER toolset is available at \url{http://fmv.jku.at/aiger/}. Accessed August 2016.} and the CUDD package for BDD manipulation (v2.5.1)~\cite{somenzi99}. 

The code is available at
\url{https://github.com/gaperez64/AbsSynthe/tree/native-dev-par}.

\subsubsection{\demiurge 1.2.0}
\demiurge was submitted by R. K\"onighofer 
and M. Seidl,
and competed in both the realizability and the synthesis track.
\demiurge implements different SAT-based methods for solving the reactive synthesis problem. In the competition, three methods are used. One approximates the winning region of the system player by repeated SAT-calls that ask for states that can enter the error states in a single step, followed by a generalization step to find additional, similar states. The second is a re-implementation of the incremental induction approach to reactive synthesis, as described by Morgenstern et al.~\cite{MorgensternGS13}. The third method encodes the reactive synthesis problem directly into a SAT problem, essentially supplying a template that fits all possible solutions, and asking the SAT solver to come up with the variable valuations that represent a winning strategy. On its own, none of the strategies can compete with the BDD-based methods. However, a combination of all three methods, were the current approximation of the winning region by the first two is used to restrict the solution template in the third approach, solved more problems than any other tool in the synthesis track of \syntcomp 2015~\cite{SYNTCOMP15}.

This year, \demiurge competed in exactly the same version as last year. For a more information on \demiurge, we refer to the original descriptions of the implemented algorithms~\cite{BloemKS14,SeidlK14}, as well as the \syntcomp 2015 report~\cite{SYNTCOMP15}. 

The code is
available at {\url{%
http://www.iaik.tugraz.at/content/research/design_verification/demiurge/}}
under the GNU Lesser General Public License version 3. 

\subsubsection{\simpleBDD}
Simple BDD Solver was submitted by L. Ryzhyk and A. Walker, and competed in the realizability track.
\simpleBDD implements the standard BDD-based algorithm for safety games, including a large number of optimizations. In particular, it includes an abstraction-refinement approach inspired by de Alfaro and Roy~\cite{dealfaro}, and is a simplified version of the solver that was
developed for the Termite project\footnote{For more information on Termite, consult \url{http://www.termite2.org}. Accessed August 2016.}~\cite{RyzhykWKLRSV14}, adapted to safety games given in the AIGER format.

It uses the BDDs with dynamic variable reordering using the sifting
algorithm~\cite{Rudell93}, and a number of additional optimizations (again, cp. Table~\ref{tab:optimizations} and the \syntcomp 2014 report~\cite{SYNTCOMP14}). 
Three configurations entered the competition:
\begin{itemize}
\item a basic algorithm without abstraction (called basic),
\item an algorithm with abstraction based on overapproximation of the winning region (called abs1), and
\item an algorithm with abstraction based on both over- and underapproximation of the winning region (called abs2).
\end{itemize}

The first two configurations are essentially identical to the configurations that entered \syntcomp 2015, and the third one is new.

\paragraph{Implementation, Availability.}
\simpleBDD is written in the Haskell functional programming language. It
uses the CUDD package for BDD manipulation (updated to v3.0.0 this year) and the
Attoparsec Haskell package for fast parsing.\footnote{Attoparsec is available at \url{https://hackage.haskell.org/package/attoparsec}. Accessed August 2016.} Altogether, the
solver, AIGER parser, compiler and command line argument parser are just over
300 lines of code. 

The code is available online at:
\url{https://github.com/adamwalker/syntcomp}.

\subsubsection{\safetysynth}
\label{sec:safetysynth}
\safetysynth was submitted by L. Tentrup, 
and competed in both the realizability and the synthesis track. \safetysynth is a re-implementation of \realizer~\cite{SYNTCOMP14,SYNTCOMP15} that implements the standard BDD-based algorithm for safety games, using the optimizations that were most beneficial for BDD-based tools in \syntcomp 2014 and 2015 (see Table~\ref{tab:optimizations}). Unlike \realizer, \safetysynth supports co-factor based strategy extraction, including . 

It competed in two configurations, which differed only in the reordering heuristics for BDDs: the \emph{basic }version uses the \texttt{GROUP\_SIFT} heuristics~\cite{PandaS95}, while the \emph{alternative }version uses \texttt{LAZY\_SIFT}. To facilitate verification of solutions, \safetysynth also produces the winning region and includes it in the solution file.

\paragraph{Implementation, Availability.}
\safetysynth is written in Swift. It uses the CUDD package (v3.0.0) for BDD manipulation, the AIGER toolset for parsing input files, and \Abc for reducing the size of solutions. 

The code is available online at: \url{https://www.react.uni-saarland.de/tools/safetysynth/}.

\subsubsection{\sdf}
\sdf was submitted by A. Khalimov, and competed in both the realizability and synthesis track. \sdf implements the standard BDD-based fixpoint algorithm, with a number of known optimizations, as well as some new heuristics.

Out of the well-known optimizations, it implements automatic reordering, eager dereferencing, direct substitution, partitioned transition relation, simultaneous conjunction and abstraction. In synthesis, it uses the co-factor-based extraction of winning strategies.

Additionally, it implements the following new heuristics:
\begin{itemize}
\item additional caching of BDDs for nodes of the specification circuit,
\item storing information about variable orders in the BDD, and limiting reordering after some time, 
\item after determining realizability, all BDDs except the non-deterministic strategy are removed, and the strategy is re-ordered once more, and
\item caching and re-use of AIGER circuits for BDD nodes during construction of the solution.
\end{itemize}

\paragraph{Implementation, Availability.}
\sdf is written in C++. It uses the CUDD package in version 3.0.0.

The code is available online at: \url{https://github.com/5nizza/sdf}.

\subsubsection{\termitesat}
\termitesat was submitted by A. Legg, N. Narodytska and L. Ryzhyk, and competed in the realizability track. \termitesat implements a novel SAT-based method for synthesis of safety specifications, as well as portfolio and hybrid modes that run the new algorithm alongside one of the algorithms of \simpleBDD.

The basic idea of the SAT-based algorithm is to explore a subset of the concrete runs of the safety game and prove that these runs can be generalized to a concrete winning strategy for one of the players. In contrast to other existing synthesis methods, it does not store or compute the set of all winning states. 
Instead, it computes an approximation of the winning states during a counter-example-guided backtracking search for candidate solutions to bounded safety games. Information about the unbounded game can be extracted in the form of Craig interpolants if a SAT query shows that a candidate strategy permits no counter example run in the bounded game. The algorithm is explained in more detail in~\cite{LeggNR16}.

For \syntcomp 2016, the algorithm has also been combined with the BDD-based algorithm with abstraction from \simpleBDD in two variants:
\begin{itemize}
\item a simple portfolio approach that runs both algorithms in parallel, and
\item a \emph{hybrid} approach that shares information between the solvers by forwarding information about states that have been determined to be winning or losing from  the SAT algorithm to the BDD algorithm. The alternate direction is not implemented so that in cases where the BDD representation of the winning states explodes, the SAT-based approach will not suffer.
\end{itemize}

\paragraph{Implementation, Availability.}
\termitesat is written in Haskell. It uses Glucose as SAT solver\footnote{Glucose is available at \url{http://www.labri.fr/perso/lsimon/glucose/}. Accessed August 2016.}~\cite{AudemardS09}, and PeRIPLO\footnote{PeRIPLO is available at \url{http://verify.inf.usi.ch/periplo}. Accessed August 2016.}~\cite{RolliniAFHS13} to find interpolants. The portfolio and hybrid modes include \simpleBDD and all its dependencies (see above).

The code is available online at: \url{https://github.com/alexlegg/TermiteSAT}.

\subsection{TLSF/LTL-Track}
\label{sec:participants-TLSF}

\subsubsection{\acaciaforaiger}
\acaciaforaiger was submitted by R. Brenguier, G. A. P\'erez, J.-F. Raskin, and O. Sankur, and competed in both the realizability and the synthesis track. \acaciaforaiger is an extension of the reactive synthesis tool Acacia+\footnote{Acacia+ is available at \url{http://lit2.ulb.ac.be/acaciaplus/}. Accessed August 2016.}, which solves the reactive synthesis problem for LTL specifications by a reduction to safety games, which can then be solved efficiently by symbolic incremental algorithms based on an \emph{antichain} representation of sets of states. Additionally, it uses a \emph{compositional} approach to solve conjuncts of a specification separately. Acacia+ has been described in more detail by Bohy et al.~\cite{bbfjr12}.

For \syntcomp 2016, Acacia+ has been extended with 
\begin{itemize}
\item a translator from TLSF to its native input format, including the separation of sub-formulas into ``spec units'' that allows for compositional solving. This translator has been integrated into the SyFCo tool. 
\item an encoding of the Mealy machines constructed during synthesis into AIGER circuits that conform to the \syntcomp rules. This encoding uses the Speculoos tools\footnote{The Speculoos tools are available at \url{https://github.com/romainbrenguier/Speculoos}. Accessed August 2016.} to parse the native Mealy machine output and generate a transducer in AIGER format.
\end{itemize}

\paragraph{Implementation, Availability.}
\acaciaforaiger is written in Python and C. It uses LTL2BA\footnote{LTL2BA is available at \url{http://www.lsv.ens-cachan.fr/~gastin/ltl2ba}. Accessed August 2016.}~\cite{go01} to convert LTL specifications into B\"uchi automata. It also includes the Speculoos and SyFco tools, as mentioned above.

The code is available online at: \url{https://github.com/gaperez64/acacia4aiger}.

\subsubsection{\bosy}
\bosy was submitted by L. Tentrup, and competed in both the realizability and the synthesis track. \bosy uses the \emph{bounded synthesis} approach~\cite{Finkbeiner13} that solves the LTL synthesis problem by first translating the specification into a universal co-B\"uchi automaton, and then encoding acceptance of a system (with bounded number of states, but unspecified behavior) into a constraint system. In contrast to the originally proposed encoding into satisfiability modulo theories (SMT), \bosy uses an encoding into quantified Boolean formulas (QBF), as described by Faymonville et al.~\cite{FaymonvilleFRT15}. One advantage of this encoding is that it allows to keep the input valuations symbolic, and to solve the constraints without explicitly enumerating all possibilities (as existing SMT solvers do). 
To detect unrealizability, the existence of a bounded strategy of the environment to falsify the specification is encoded in a similar way, and checked in parallel.

The resulting QBF formulas have the quantifier structure $\exists \forall \exists$. 
Before solving them, they are first simplified, using the QBF preprocessor bloqqer\footnote{Bloqqer is available at \url{http://fmv.jku.at/bloqqer}. Accessed August 2016.}. In realizability mode, the simplified formula is then directly solved, using the QBF solver RAReQS~\cite{JanotaKMC16}. In synthesis mode, \bosy uses RAReQS only to get a satisfying assignment for the outer existential quantifier, and reducing the query to a 2QBF formula with quantifier structure $\forall \exists$. This 2QBF formula is then solved by the certifying QBF solver QuAbS~\cite{Tentrup16}, and the returned certificate represents a solution to the synthesis problem. This solution is then converted into AIGER format, and further simplified using the \Abc framework. 

Two configurations of \bosy competed in \syntcomp 2016, differing in how the bound for the implementation size is increased after finding that no solution exists for the current bound: one configuration increases the bound linearly, the other exponentially. \bosy supports both Mealy and Moore semantics natively, as well as the extraction of a winning strategy for the environment in case of unrealizable specifications.

\paragraph{Implementation, Availability.}
\bosy is written in Swift. It uses LTL3BA\footnote{LTL3BA is available at \url{https://sourceforge.net/projects/ltl3ba/}. Accessed August 2016.}~\cite{ltl3ba} to convert LTL specifications into B\"uchi automata. It uses bloqqer, RAReQS\footnote{RAReQS is available at \url{http://sat.inesc-id.pt/~mikolas/sw/areqs/}. Accessed August 2016.} and QuAbs\footnote{QuAbs \available{https://www.react.uni-saarland.de/tools/quabs/}} to solve QBF constraints, as mentioned above. Finally, it uses \syfco to translate TLSF specifications to its native input format, and \Abc to simplify solutions.

The code is available online at: \url{https://www.react.uni-saarland.de/tools/bosy/}.

\subsubsection{\party}
\party was submitted by A. Khalimov, and competed in both the realizability and the synthesis track. \party also uses the \emph{bounded synthesis} approach~\cite{Finkbeiner13} for solving LTL synthesis problems. It uses an encoding into SMT formulas, as described in the original approach. To detect unrealizability, it also checks for the existence of a strategy for the environment to falsify the specification, but only for very simple solutions, i.e., stateless strategies. The tool at the algorithms it implements have been described in more detail by Khalimov et al.~\cite{KhalimovJB13,KhalimovJB13a}.

To solve the synthesis problem more efficiently, \party uses the following optimizations of the original approach:
\begin{itemize}
\item it uses a heuristic \emph{strengthening} of assume-guarantee specifications: if the specification has both liveness and safety assumptions, then it first tries to synthesize a solution where safety guarantees have to hold whenever the safety assumptions hold, regardless of the liveness assumptions; should no solution exist, then it falls back to the full formula;
\item to avoid duplication of work when increasing the bound, the SMT solver is used in incremental mode: whenever the bound increases, the constraints that are specific to the previous bound are retracted, and constraints for the new bound are added,
\item an optimization that analyses the strongly connected components (SCCs) of the automaton that is obtained from the specification, and simplifies the encoding if there are multiple SCCs.
\end{itemize}

When the SMT solver finds a satisfying assignment to the constraints, the solution is first encoded into Verilog, and then converted to AIGER and simplified using existing tools.

\paragraph{Implementation, Availability.}
\party is written in Python. It uses LTL3BA to convert LTL specifications into B\"uchi automata. It uses Z3\footnote{Z3 is available at \url{https://github.com/Z3Prover/z3}. Accessed August 2016.}~\cite{Moura08} for solving SMT constraints. For converting generated Verilog code into AIGER, it uses \texttt{vl2mv} (contained in VIS\footnote{VIS is available at \url{http://vlsi.colorado.edu/~vis/}). Accessed August 2016.}~\cite{VIS} and \Abc. Finally, it uses \syfco to translate TLSF specifications to its native input format.

The code is available online at: \url{https://github.com/5nizza/party-elli}.

%% file: results.tex
\section{Experimental Results}
\label{sec:results}

We present the results of \syntcomp 2016, separated into the AIGER/safety-track and the TLSF/LTL-track. Both main tracks are separated into realizability and synthesis subtracks, and parallel and sequential execution modes.
Detailed results of the competition are also directly accessible via the web-frontend of our instance of the EDACC platform at \url{http://syntcomp.cs.uni-saarland.de}.

\subsection{AIGER/safety-Track: Realizability}

In the track for realizability checking of safety specifications in AIGER format, $6$ tools competed on $234$ benchmark instances, selected from the different benchmark categories as explained in Section~\ref{sec:selection}. Overall, $17$ different configurations entered this track, with $11$ using sequential execution mode and $6$ using parallel mode. In the following, we compare the results of these $17$ configurations on the $234$ benchmarks selected for \syntcomp 2016. To assess improvements compared to previous competitions, we also give results of $4$ of the best-performing tool configurations from \syntcomp $2014$ and $2015$. 

We first restrict the evaluation of results to purely sequential tools, then extend it to include also the parallel versions, and finally give a brief analysis of the results.

\paragraph{Sequential Mode.}
In sequential mode, \abssynthe competed with three configurations (seq1, seq2 and seq3), \demiurge with one configuration (D1real), \simpleBDD with three configurations (basic, abs1, abs2), \safetysynth with two configurations (basic and alternative), as well as \sdf and \termitesat with one configuration each.

The number of solved instances per configuration, as well as the number of uniquely solved instances, are given in Table~\ref{tab:results-realseq}. No tool could solve more than $175$ out of the $234$ instances, or about $75\%$ of the benchmark set. $20$ instances could not be solved by any tool within the timeout. 
For comparison, we also ran a number of additional tools on the benchmark set: the best-performing sequential configuration of \abssynthe from \syntcomp 2015, as well as the two winning configurations from \syntcomp 2014 and 2015, both from \simpleBDD.\footnote{We do not consider last year's version of \demiurge separately, since it participates in the same version as last year.}
The results for these tools are grayed out in the table, and are not counted towards unique solutions. 

\begin{table}[h]
\caption{Results: AIGER Realizability (sequential mode only)}
\label{tab:results-realseq}
\centering
\def\arraystretch{1.3}
{\sffamily \small
\begin{tabular}{@{}llll@{}}
\toprule
Tool & (configuration)		& Solved  &  Unique \\ 
\midrule
\simpleBDD & (abs1) & 175 		& 1 \\
\simpleBDD & (abs2) & 167     & 1 \\
\safetysynth & (basic)      & 164     & 0 \\
\simpleBDD & (basic)        & 164     & 0 \\
\safetysynth & (alternative) & 163    & 0 \\
\grayed{\simpleBDD} & \grayed{(2014)}         & \grayed{163} \\
\grayed{\simpleBDD} & \grayed{(2015, 2)}			& \grayed{163} \\	
\abssynthe & (seq3) 	 	    & 159     & 4 \\
\grayed{\abssynthe} & \grayed{(2015, seq2)} 	& \grayed{159} \\
\abssynthe & (seq2) 	 	    & 149     & 0 \\
\sdf 			& 								& 149			& 0 \\
\abssynthe & (seq1)					& 145			& 0 \\
\demiurge & (D1real)				& 129			& 6 \\
\termitesat	& 							& 97			& 4 \\
\bottomrule
\end{tabular}
}
\end{table}

\begin{figure}
\centering
\includegraphics[width=1\linewidth]{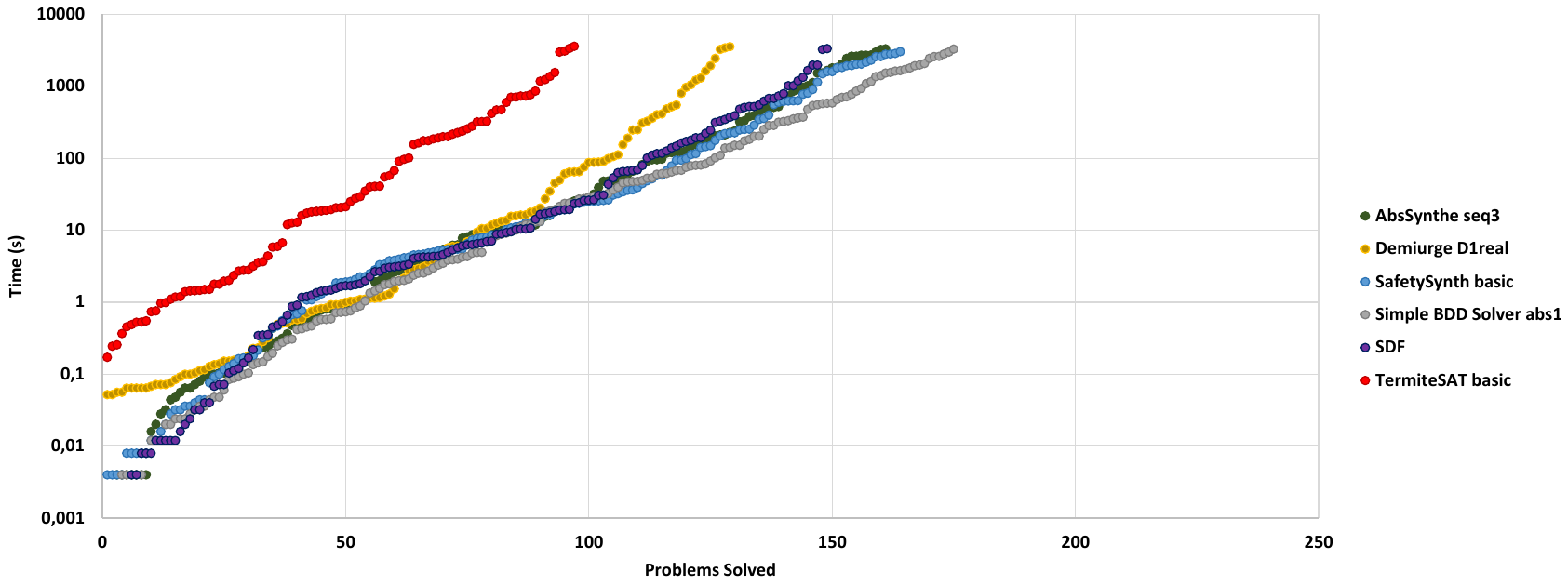}
\caption{Runtime Cactus Plot of Best Sequential Configurations}
\label{fig:cactus-realseq}
\end{figure}

Figure~\ref{fig:cactus-realseq} gives a cactus plot for runtimes of the best-performing sequential configuration of each tool.
%
%


\begin{figure}
\centering
\includegraphics[width=1\linewidth]{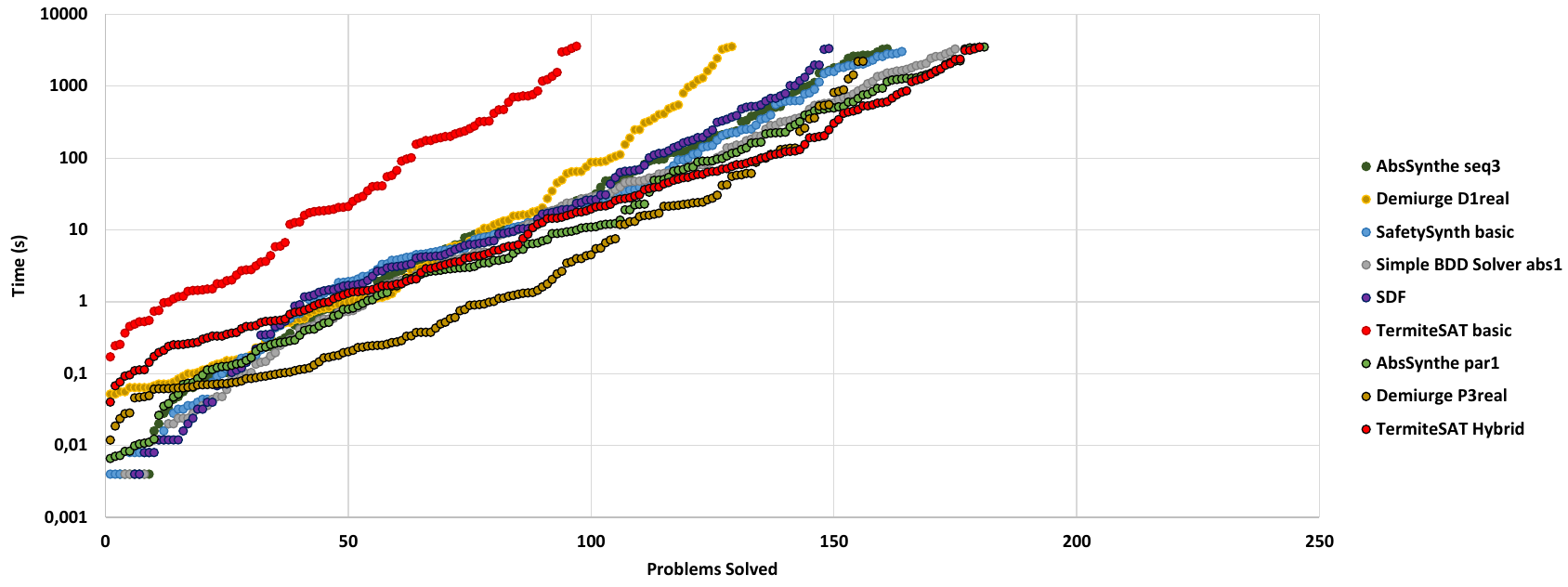}
\caption{Runtime Cactus Plot of Best Configurations Overall}
\label{fig:cactus-realall}
\end{figure}	

\paragraph{Parallel Mode.}
Three of the tools that entered the competition had at least one parallel configuration for the realizability track: three configurations of \abssynthe (par1, par2, par3), one configuration of \demiurge (P3real), and two configurations of \termitesat (portfolio, hybrid). The parallel configurations had to solve the same set of benchmark instances as in the sequential mode. In contrast to the sequential mode, runtime of tools is now measured in wall time instead of CPU time. The results are given in Table~\ref{tab:results-realpar}. Compared to sequential mode, a number of additional instances could be solved: both \abssynthe and \termitesat have one or more configurations that solve more then the best tool in sequential mode (about $77\%$ of the benchmark set).
Only $14$ instances could not be solved by any tool in either sequential or parallel mode.

\begin{table}
\caption{Results: AIGER Realizability (parallel mode only)}
\label{tab:results-realpar}
\centering
\def\arraystretch{1.3}
{\sffamily \small
\begin{tabular}{@{}llll@{}}
\toprule
Tool & (configuration)			& Solved  &  Unique \\ 
 \midrule
\abssynthe & (par1) 	 	 	& 181 & 1 \\
\termitesat & (hybrid) 		& 180 & 0 \\
\termitesat & (portfolio) & 179 & 0 \\
\grayed{\abssynthe} & \grayed{(2015, par1)} & \grayed{172}\\
\demiurge & (P3real) 			& 156 & 5 \\
\abssynthe & (par3) 			& 148	& 0 \\
\abssynthe & (par2)				& 141 & 0 \\
\bottomrule
\end{tabular}
}
\end{table}

Note that in Table~\ref{tab:results-realpar} we only count a benchmark instance as uniquely solved if it is not solved by any other configuration, including the sequential configurations. Still, \abssynthe (par1) can provide one unique solution, and \demiurge (P3real) five.

We also ran this year's benchmark selection on the best parallel configuration from \syntcomp 2015. The results fo \abssynthe (2015, par1) appear grayed out in the table. This was the only parallel configuration that we ran from previous versions, since \abssynthe was the only participating tool with a parallel mode that ran in \syntcomp 2015 and was not the same as one of the configurations that ran this year, and in \syntcomp 2014 the best configuration overall was a sequential configuration. 

\begin{figure}[h]
\centering
\includegraphics[width=\linewidth]{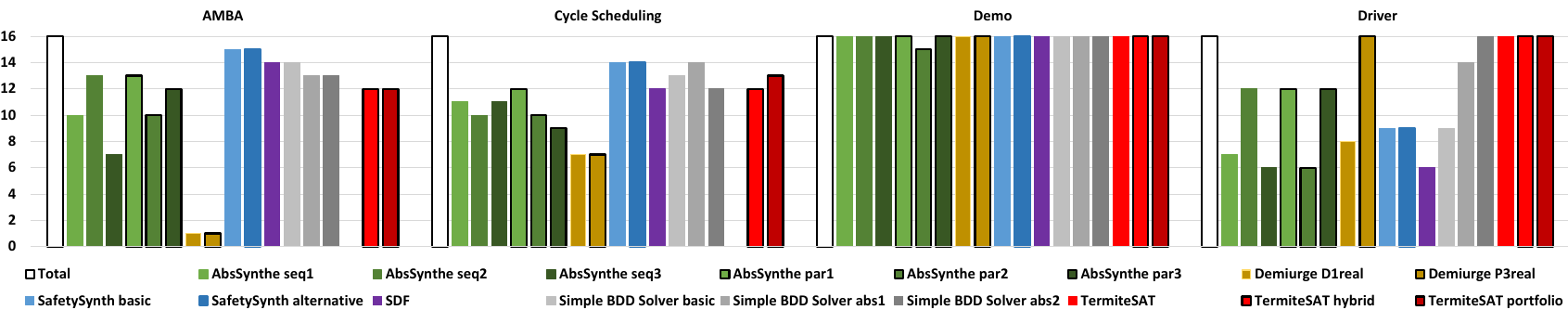}
%
\vspace{1em}

\includegraphics[width=\linewidth]{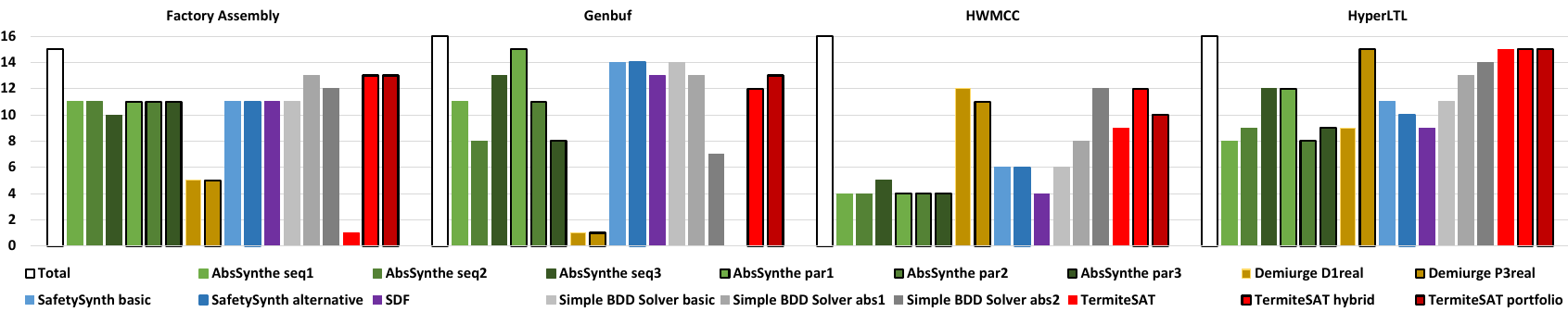}
\caption{AIGER/safety Realizability Track, Solved Instances by Category (part 1)}
\label{fig:bycat2}
\end{figure}

\begin{figure}[h]
\centering
\includegraphics[width=\linewidth]{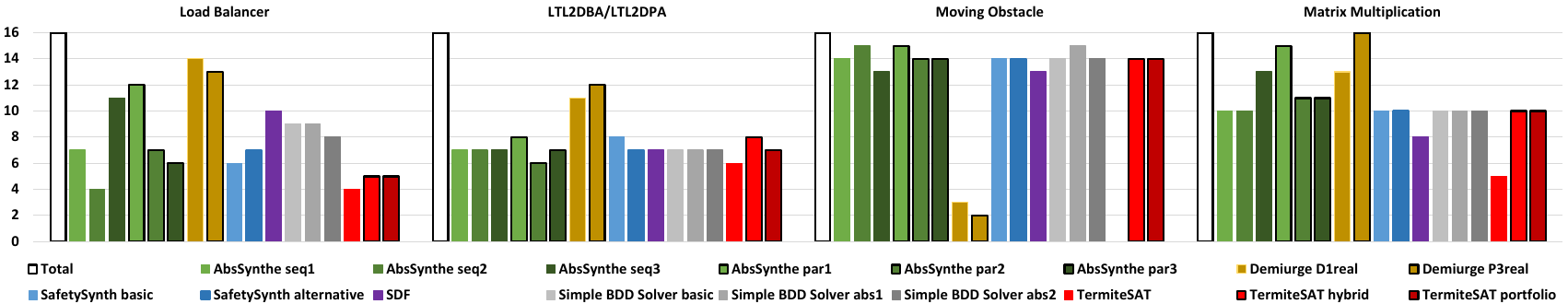}
%
\vspace{1em}

\includegraphics[width=\linewidth]{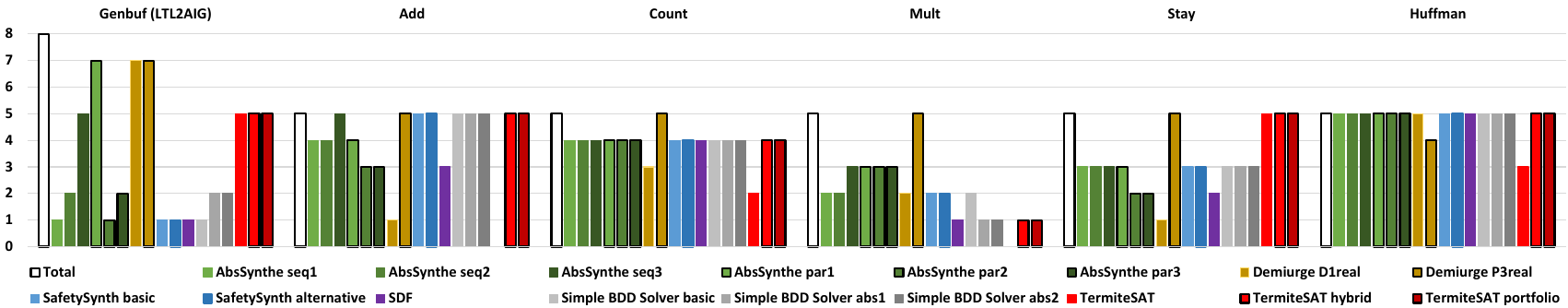}
\caption{AIGER/safety Realizability Track, Solved Instances by Category (part 2)}
\label{fig:bycat4}
\end{figure}

\paragraph{Both modes: Solved Instances by Category.} 
For both the sequential and the parallel configurations, Figures~\ref{fig:bycat2} and \ref{fig:bycat4} give an overview of the number of solved instances per configuration and category (as defined in Table~\ref{tab:selected-benchmarks}).


\paragraph{Analysis.}
We note that on this year's benchmark set, \simpleBDD (abs1) can solve significantly more problems than its best configuration from last year. Since these two configurations use the same algorithm, the main difference between them is the update to CUDD v3.0.0, which seems to give a significant performance boost. The new algorithm \simpleBDD (abs2) that uses also an underapproximation of the winning states cannot solve the same number of problems, but does solve a number of problems that \simpleBDD (abs1) does not solve, including one unique solution among all sequential configurations. 

\safetysynth solves fewer problems than \simpleBDD (abs1 and abs2), but its basic configuration still solves one additional instance, compared to last year's winner, \simpleBDD (2015, 2). The alternative version with a different BDD reordering scheme shows a very similar behavior, and solves one instance less overall. Note that \safetysynth is also based on CUDD v3.0.0, so it presumably benefits from the same perfomance boost as \simpleBDD, compared to tools that use an earlier version.

The best sequential configuration of \abssynthe solves exactly the same number of problems as its best configuration last year, and less than both \simpleBDD and \safetysynth. 
Comparing the different sequential configurations, we note that they have similar behavior on many categories, but are also substantially different on some, e.g. AMBA, Driver, and Genbuf (LTL2AIG) (cp. Figure~\ref{fig:bycat4}). Overall, the configuration seq3 that uses both compositionality and abstraction solves the highest number of problems, and the configuration seq1 that uses neither solves the least number. 
The parallel configuration \abssynthe (par1) solves the highest number of problems among all configurations, and shows that a well-chosen portfolio approach can solve significantly more problems than any single algorithm. Note also that \abssynthe still uses CUDD in version 2.5.1, so it could possibly benefit additionally from an upgrade to the newer version. The parallel configurations par2 and par3, which represent portfolios of seq1 and seq2, respectively, that only differ in the BDD reordering, solve fewer instances than their sequential counterparts.

We note that the new participant \sdf solves the same number of problems as \abssynthe (seq2). Finally, the SAT-based algorithms of \demiurge and \termitesat are in general not competitive in sequential mode, but each of them solves a significant number of problems uniquely. That is, they solve problems that none of the BDD-based algorithms can solve, but that are also not shared between the two SAT-based algorithms. In their parallel configurations, both tools show their strengths: \termitesat (hybrid) solves only one problem less than the best configuration overall, and \demiurge (P3real) solves a good number of instances, while supplying $5$ solutions to problems that no other (sequential or parallel) configuration could solve. These additional problems can be attributed to the additional algorithms that are used in configuration P3real, including exchange of information between different algorithms. In contrast, the hybrid mode of \termitesat that also communicates information between algorithms did not produce unique solutions, and only solved one additional problem compared to the portfolio configuration without communication. 

\subsection{AIGER/safety-Track: Synthesis}
In the track for synthesis from safety specifications in AIGER format, participants had to solve the same set of benchmarks as in the realizability track, except that we removed instances that could not be solved by at least one configuration in the realizability track. Thus, the benchmark set for the synthesis track contains $215$ instances.
Four tools entered the track: \abssynthe with three sequential and three parallel configurations, \demiurge with one configuration for each mode, \safetysynth with two sequential configurations, and \sdf with one sequential configuration.

For \syntcomp 2016, the ranking in the synthesis track is only based on the number of instances that can be solved within the timeout. The difference to the realizability track is that a solution for a realizable specification is only considered as correct if it can be model-checked within a separate timeout of one hour (cf. Section~\ref{sec:setup}).
As before, we start by presenting the results for the sequential configurations, followed by parallel configurations, and end with an analysis of the results.

\paragraph{Sequential Mode.}
In this mode, \abssynthe competed with three configurations (seq1, seq2, seq3), \demiurge with one configuration (D1synt), \safetysynth with two configurations (basic, alternative), and \sdf with one configuration. 

Table~\ref{tab:results-syntseq} summarizes the experimental results, including the number of solved benchmarks, the uniquely solved instances, and the number of solutions that could not be model-checked within the timeout. Note that the ``solved'' column gives the number of problems that have either been correctly determined unrealizable, or for which the tool has presented a solution that could be verified. With this requirement, no sequential configuration could solve more than $153$ or about $71\%$ of the benchmarks, and $23$ instances could not be solved by any tool. \sdf is the only tool that generates a significant number of solutions that could not be verified.


\begin{table}[h]
\caption{Results: AIGER Synthesis (sequential mode only)}
\label{tab:results-syntseq}
\centering
\def\arraystretch{1.3}
{\sffamily \small
\begin{tabular}{@{}lllll@{}}
\toprule
Tool & (configuration) & Solved & Unique & MC Timeout\\
\midrule
\safetysynth & (basic) 	& 153 & 0 & 0\\
\safetysynth & (alt)		& 152 & 0 & 0\\
\abssynthe 	& (seq3) 		& 151 & 2 & 1\\
\abssynthe 	& (seq2) 		& 140 & 0 & 0\\
\abssynthe 	& (seq1) 		& 137 & 0 & 0\\
\sdf 				& 					& 134 & 0 & 11\\
\demiurge 	& (D1synt) 	& 118 & 2 & 3\\
\bottomrule
\end{tabular}
}
\end{table}

\paragraph{Parallel Mode.}
In this mode, \abssynthe competed with three configurations (par1, par2, par3), and \demiurge with one configuration (P3synt). 

Table~\ref{tab:results-syntpar} summarizes the experimental results, again including the number of solved benchmarks, the uniquely solved instances, and the number of solutions that could not be verified within the timeout. No tool solved more than $165$ problem instances, or about $77\%$ of the benchmark set. There are only very few (potential) solutions that could not be verified within the timeout.
Like in the parallel realizability track, we only consider instances as uniquely solved if they are not solved by any other configuration, including sequential ones. Still, \demiurge (P3Synt) was able to supply $9$ unique solutions.


\begin{table}[h]
\caption{Results: AIGER Synthesis (parallel mode only)}
\label{tab:results-syntpar}
\centering
\def\arraystretch{1.3}
{\sffamily \small
\begin{tabular}{@{}lllll@{}}
\toprule
Tool & (configuration) & Solved & Unique & MC Timeout\\
\midrule
\abssynthe  & (par1) 		& 165 & 0 & 0\\
\demiurge 	& (P3Synt) 	& 150 & 9 & 0\\
\abssynthe 	& (par3) 		& 140 & 0 & 1\\
\abssynthe 	& (par3) 		& 134 & 0 & 1\\
\bottomrule
\end{tabular}
}
\end{table}
\paragraph{Analysis.}
Unsurprisingly, the number of solved instances for each tool in the synthesis track corresponds roughly to those solved in the realizability track. With respect to the smaller number of participants, both \abssynthe (seq3) and \demiurge (D1synt) provide two unique solutions in the sequential mode, and \demiurge (D1synt) provides an additional $9$ solutions that are unique over all sequential and parallel configurations. 

As mentioned, \sdf is the only tool that produces a significant number of solutions that cannot be verified. This is due to the fact that \demiurge usually produces small solutions, and both \abssynthe and \safetysynth use the possibility to also produce a winning region that is used for verification. Thus, we can conclude that the introduction of additional witness information was successful and almost completely solves the problem of verification.\footnote{This is at least true in the current setup, where we first try a simple invariant check of the solution with respect to the winning region, and fall back to model checking if this check fails or times out.}

\begin{figure}[h]
\centering
\includegraphics[width=\linewidth]{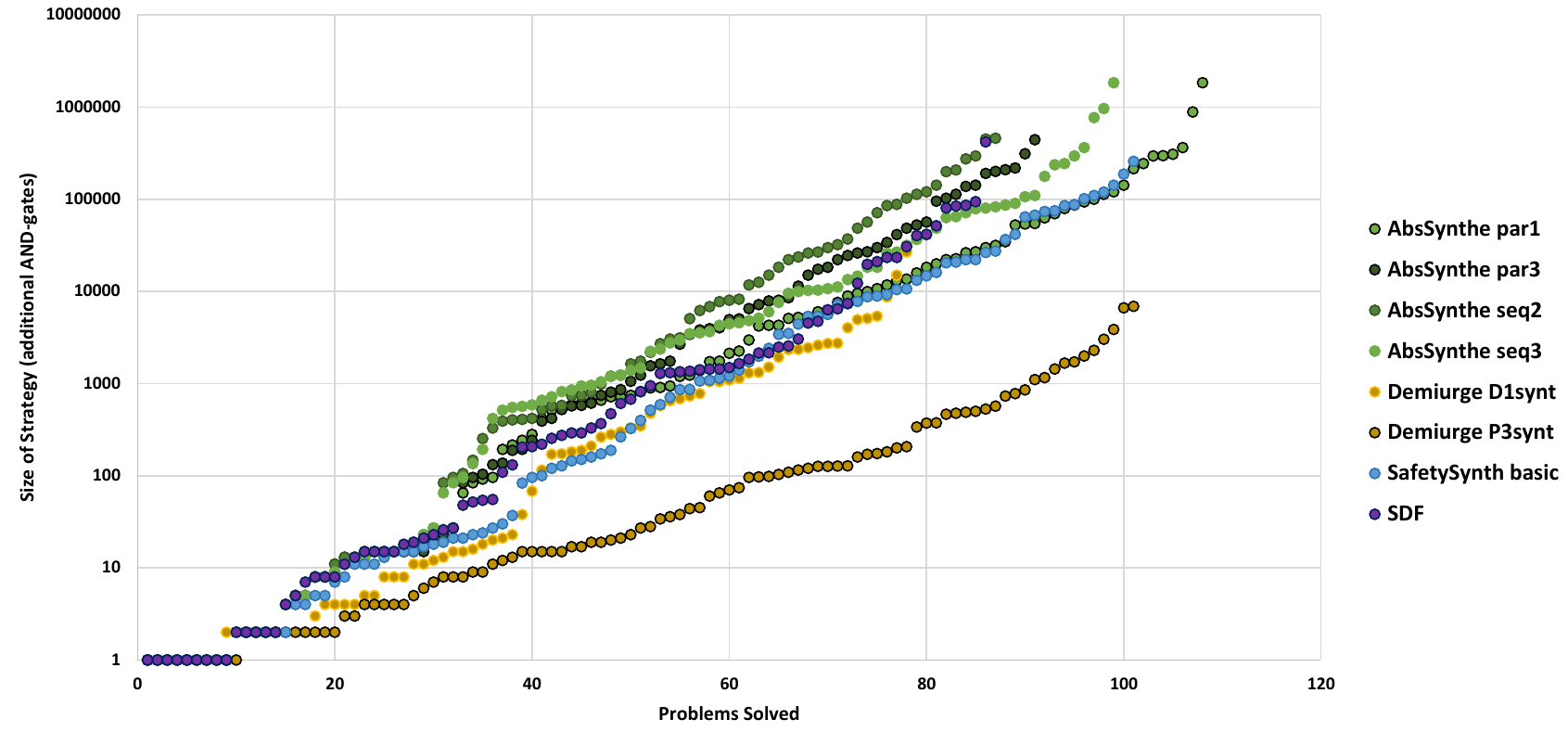}
\caption{AIGER/safety Synthesis Track: Size of Solution Strategies for Selected Configurations}
\label{fig:size-cactus}
\end{figure}

Even though for \syntcomp 2016 we do not have a ranking that is based on the size of solutions, we want to give a quick comparison. Figure~\ref{fig:size-cactus} plots the sizes of synthesized strategies for some of the configurations. In contrast to previous size comparisons, we consider here not the size of the complete solution (which includes the specification circuit), but only the number of additional AND-gates, which corresponds to the strategy of the controller. Note that the solutions of \demiurge (P3synt) are often more than an order of magnitude smaller than the smallest BDD-based solutions, and there is sometimes another order of magnitude between the smallest and largest of those.

%% file: results-tlsf.tex
\subsection{TLSF/LTL-Track: Realizability}

In the track for realizability checking of LTL specifications in TLSF, $3$ tools competed on $195$ benchmark instances, selected as explained in Section~\ref{sec:selection}. The three tools competed in $4$ sequential and $2$ parallel configurations. In the following, we compare the results of these $6$ configurations on the $195$ benchmarks selected for \syntcomp 2016. Additionally, we give a comparison to the legacy synthesis tool \unbeast\footnote{\unbeast is available at \url{http://www.react.uni-saarland.de/tools/unbeast/}. Accessed August 2016.}~\cite{Ehlers11}, which was not entered as a participant, but only modified by the organizers in order to allow a rough comparison to the actual participants. 

Again, we first restrict the evaluation of results to sequential configurations, then extend it to include parallel configurations, and finally give a brief analysis.

\paragraph{Sequential Mode.}
In sequential mode, \bosy competed with two configurations (lin and exp), and \acaciaforaiger and \party competed with one configuration each. 

The number of solved instances per configuration, as well as the number of uniquely solved instances, are given in Table~\ref{tab:results-realseq-tlsf}. No tool could solve more than $153$ out of the $195$ instances, or about $78\%$ of the benchmark set. $10$ instances could not be solved by none of the participants within the timeout. Additional results for legacy tool \unbeast are grayed out in the table, and are not counted towards solved instances or unique solutions. 

\begin{table}[h]
\caption{Results: TLSF Realizability (sequential mode only)}
\label{tab:results-realseq-tlsf}
\centering
\def\arraystretch{1.3}
{\sffamily \small
\begin{tabular}{@{}llll@{}}
\toprule
Tool & (configuration)		& Solved  &  Unique \\ 
\midrule
\acaciaforaiger	& 			& 153	&	34\\
\bosy 					& (exp) & 147	& 1\\
\bosy						& (lin) & 138	& 0\\
\party					&				& 118	& 0\\
\grayed{\unbeast}	&				& \grayed{65}	& \grayed{3}\\
\bottomrule
\end{tabular}
}
\end{table}

\begin{figure}
\centering
\includegraphics[width=1\linewidth]{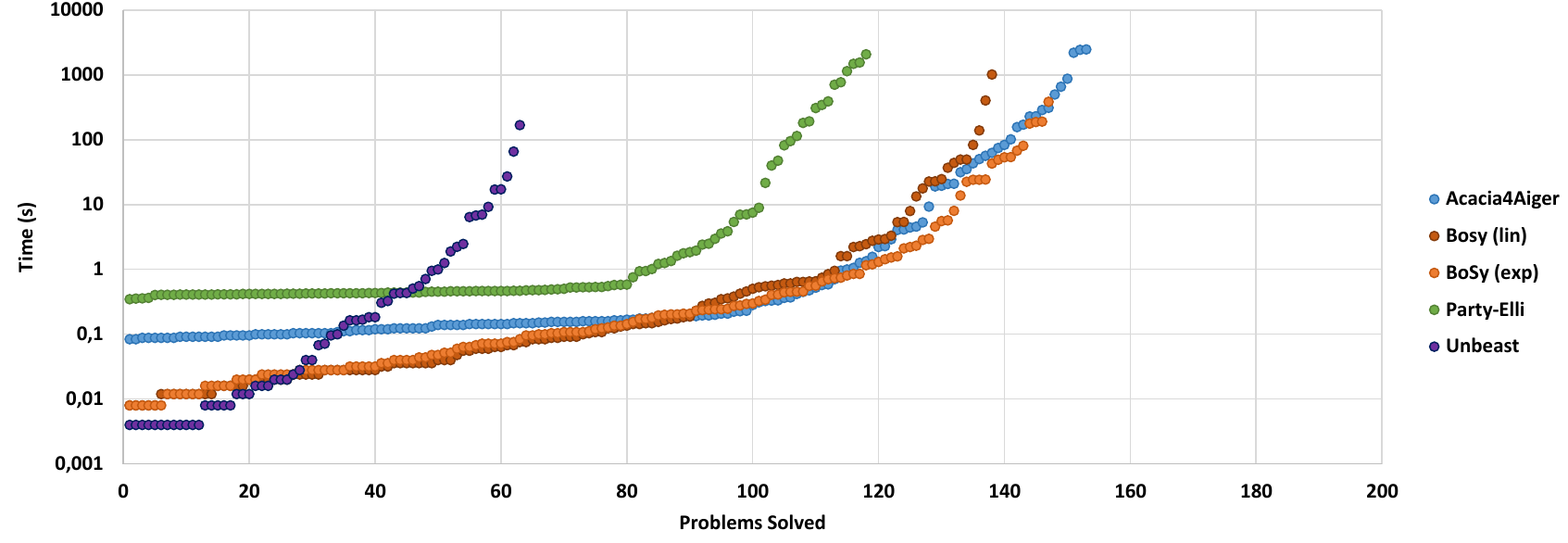}
\caption{TLSF/LTL Realizability Track: Runtimes of Sequential Configurations}
\label{fig:cactus-realseq-tlsf}
\end{figure}

Figure~\ref{fig:cactus-realseq-tlsf} gives a cactus plot for runtimes of all sequential algorithms in the realizability track.
%
%


\begin{figure}
\centering
\includegraphics[width=1\linewidth]{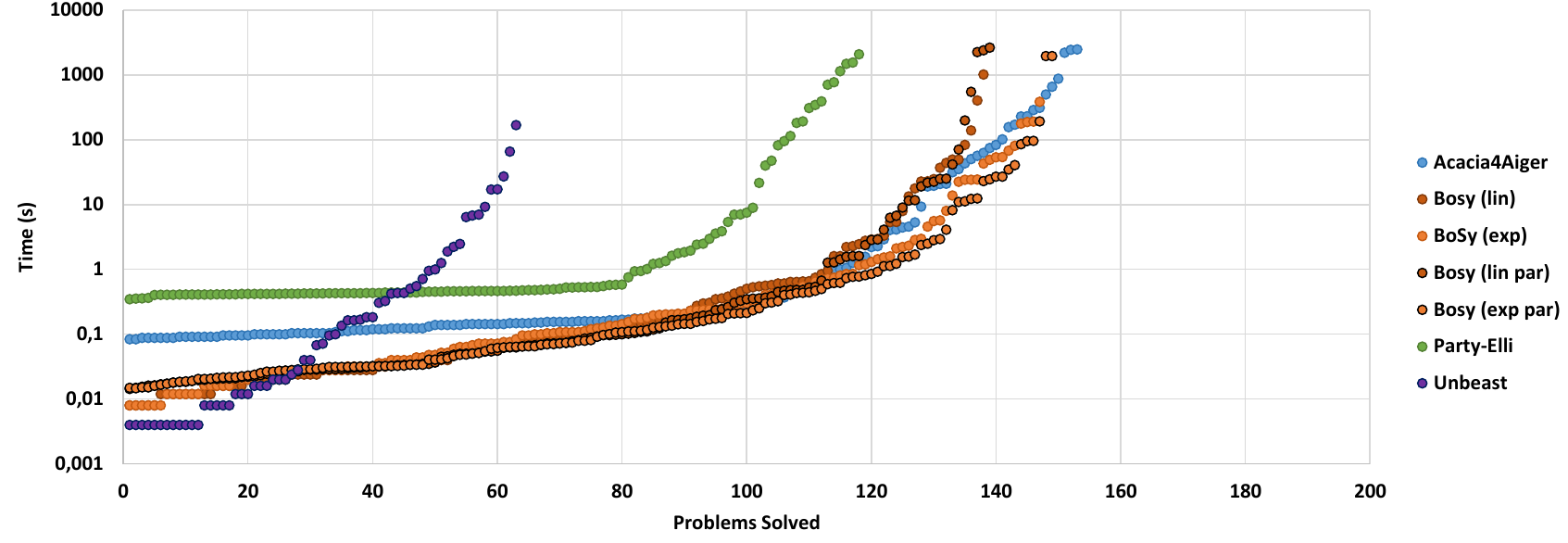}
\caption{TLSF/LTL Realizability Track: Runtimes of All Configurations}
\label{fig:cactus-realall}
\end{figure}	

\paragraph{Parallel Mode.}
The only tool that also competed in parallel configurations was \bosy, with configurations lin par and exp par. As before, parallel configurations solve the same set of benchmark instances as in the sequential mode, but runtime is measured in wall time instead of CPU time. The results are given in Table~\ref{tab:results-realpar-tlsf}. The parallel configurations of \bosy only solve a small number of additional instances. 

\begin{table}
\caption{Results: TLSF Realizability (parallel mode only)}
\label{tab:results-realpar-tlsf}
\centering
\def\arraystretch{1.3}
{\sffamily \small
\begin{tabular}{@{}llll@{}}
\toprule
Tool & (configuration)			& Solved  &  Unique \\ 
 \midrule
\bosy 		& (exp par)			& 149	& 0\\
\bosy			& (lin par)			& 139	& 0\\
\bottomrule
\end{tabular}
}
\end{table}

Since in Table~\ref{tab:results-realpar-tlsf} we again only count a benchmark instance as uniquely solved if it is solved by neither another sequential nor another parallel configuration, there are no unique solutions.

\begin{figure}[h]
\centering
\includegraphics[width=.7\linewidth]{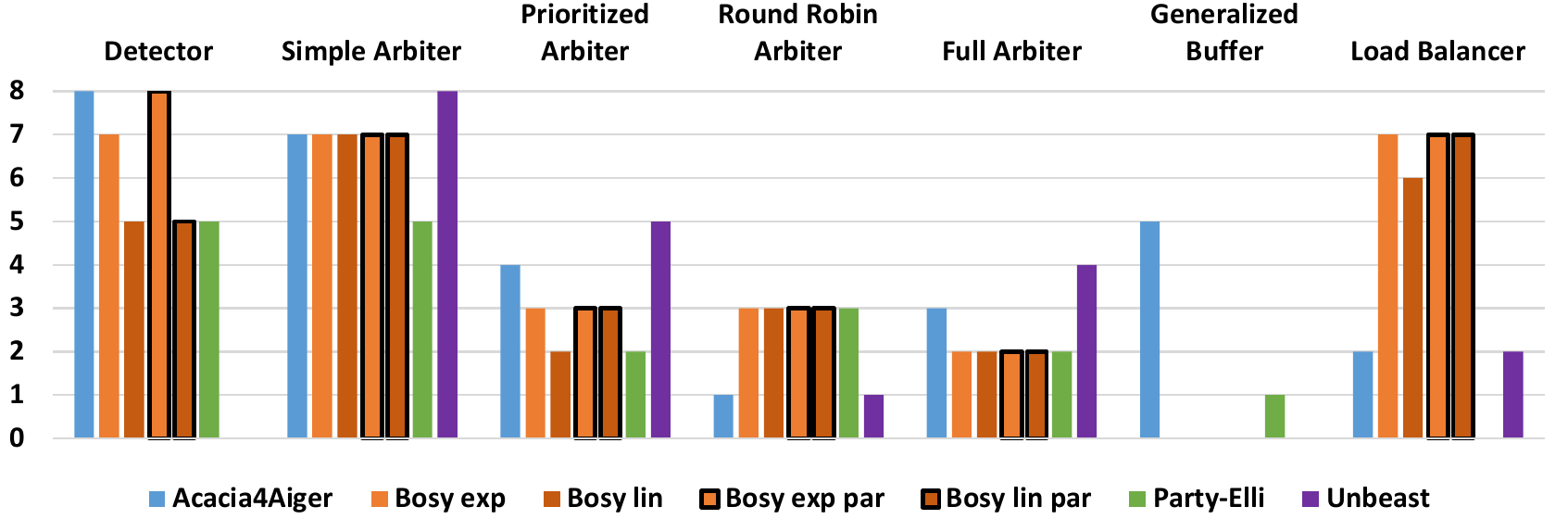}
%
\vspace{1em}

\includegraphics[width=.7\linewidth]{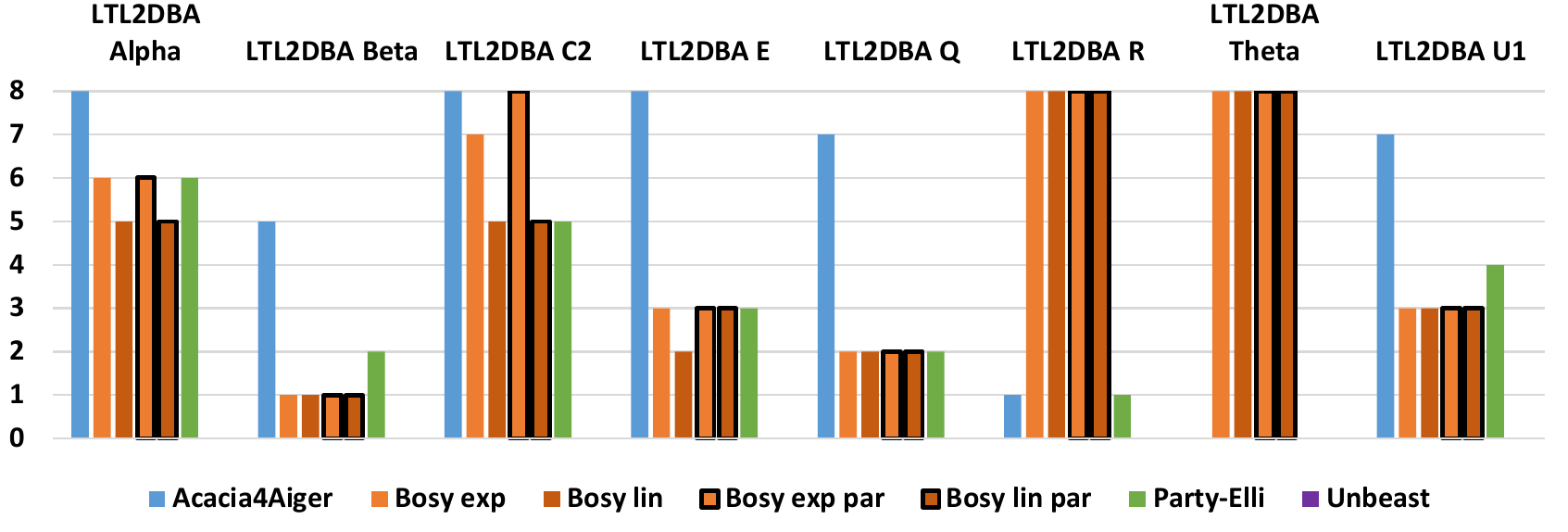}
\caption{TLSF/LTL Realizability Track: Solved Instances for Parameterized Benchmarks}
\label{fig:bycat-tlsf}
\end{figure}

\paragraph{Both modes: Solved Instances of Parameterized Benchmarks.} 
For both the sequential and the parallel configurations, Figure~\ref{fig:bycat-tlsf} gives an overview of the number of solved instances per configuration and parameterized benchmark.

\paragraph{Analysis.}
We note that \acaciaforaiger solves slightly more instances than \bosy (exp par), the most successful configuration that is based on a bounded synthesis algorithm. Since \acaciaforaiger has a high number of unique solutions, there must also be a high number of instances that is only solved by 2 or more configurations of the bounded synthesis tools. Thus, we conclude that the different algorithms have significantly different strengths. This also shows in the analysis of the parameterized benchmarks in Figure~\ref{fig:bycat-tlsf}, where \acaciaforaiger clearly dominates for the benchmarks generalized buffer and the LTL2DBA benchmarks $\alpha, \beta$, E, Q, and U1, while the bounded synthesis-based tools dominate on the round robin arbiter, the load balancer, and LTL2DBA benchmarks R and $\theta$. 

A remarkable result is that \acaciaforaiger has not solved a single unrealizable benchmark. This did not keep it from solving the highest number of instances, since only $20$ of the $195$ available benchmarks were unrealizable. Thus, a set of benchmarks that contained more unrealizable instances might have lead to a very different result. 

Also, \party did not solve a single instance of the load balancer benchmarks. A deeper analysis shows that this is due to the strategy applied in the strengthening optimization (see the description in Section~\ref{sec:participants-TLSF}): the strengthened formula is unrealizable, and the algorithm tries to find ever larger solutions for this unrealizable formula.

The parallel versions of \bosy were a bit more successful than the sequential versions, but the number of additionally solved instances is small.

Finally, legacy tool \unbeast was not very successful overall, but did provide some unique solutions and solved the highest number of instances for some parameterized examples, namely the simple arbiter, prioritized arbiter, and full arbiter benchmarks.

\subsection{TLSF/LTL-Track: Synthesis}
In the track for synthesis from LTL specifications in TLSF, participants had to solve the same benchmarks as in the LTL/TLSF-realizability track, except that we removed instances that remained unsolved in the realizability track. Thus, the benchmark set for the synthesis track contains $185$ instances.
The track had the same participants as the LTL/TLSF-realizability track: \acaciaforaiger with one configuration, \bosy with four configurations, and \party with one configuration. Legacy tool \unbeast did not run in this track, since it would have been an additional implementation effort to encode its native output into AIGER.

As for the AIGER/safety-track, the ranking is only based on the number of instances that can be solved within the timeout, and a solution for a realizable specification is only considered correct if it can be model-checked within a separate timeout of one hour (cf. Section~\ref{sec:setup}).
Again, we start by presenting the results for the sequential configurations, followed by parallel configurations, and end with an analysis of the results.

\paragraph{Sequential Mode.}
Table~\ref{tab:results-syntseq-tlsf} summarizes the experimental results for the sequential configurations, including the number of solved benchmarks, the uniquely solved instances, and the number of solutions that could not be model-checked within the timeout. 

As before, the ``solved'' column gives the number of problems that have either been correctly determined unrealizable, or for which the tool has presented a solution that could be verified. With this requirement, no sequential configuration could solve more than $136$ or about $74\%$ of the benchmarks, and $22$ instances could not be solved by any tool. 

In this track, \acaciaforaiger is the only tool that generates solutions that could not be verified. Moreover, the submitted version of \acaciaforaiger produced $4$ solutions that were determined to be incorrect by the model checker. The bug was fixed after the competition, and the results for the fixed version are also given in the table (grayed out). The $4$ problems that are solved additionally by the fixed version are exactly those for which the originally submitted version produced a wrong solution.


\begin{table}[h]
\caption{Results: TLSF Synthesis (sequential mode only)}
\label{tab:results-syntseq-tlsf}
\centering
\def\arraystretch{1.3}
{\sffamily \small
\begin{tabular}{@{}llllll@{}}
\toprule
Tool & (configuration) & Solved & Unique & MC Timeout & Wrong Solution\\
\midrule
\bosy 					& (exp)			& 136	& 0	& 0 & 0\\
\bosy 					& (lin)			& 129 & 0 & 0 & 0\\
\party					& 					& 119 & 6 & 0 & 0\\
\hdashline
\acaciaforaiger & 					& 133	& 16 & 20 & {\bf \textcolor{red}{4}}\\
\grayed{\acaciaforaiger} & \grayed{(bugfix)} & \grayed{137} & & \grayed{20} & \grayed{0}\\
\bottomrule
\end{tabular}
}
\end{table}

\paragraph{Parallel Mode.}
In this mode, \bosy competed with two configurations (lin par, exp par). 
Table~\ref{tab:results-syntpar-tlsf} summarizes the experimental results, in the same format as before. No configuration solved more than $140$ problem instances, or about $76\%$ of the benchmark set. All solutions were verified within the timeout.
As before, we only consider instances as uniquely solved if they are not solved by any other configuration, including sequential ones. Still, \bosy (exp par) provided $3$ unique solutions.


\begin{table}[h]
\caption{Results: TLSF Synthesis (parallel mode only)}
\label{tab:results-syntpar-tlsf}
\centering
\def\arraystretch{1.3}
{\sffamily \small
\begin{tabular}{@{}llllll@{}}
\toprule
Tool & (configuration) & Solved & Unique & MC Timeout & Wrong Solution\\
\midrule
\bosy 					& (exp par)			& 140	& 3	& 0 & 0\\
\bosy 					& (lin par)			& 130 & 0 & 0 & 0\\
\bottomrule
\end{tabular}
}
\end{table}
\paragraph{Analysis.}
As for the AIGER/safety-track, the number of solved instances for each tool in synthesis is closely related to the solved instances in realizability checking. The number of unique instances of \acaciaforaiger decreases significantly, in part due to the high number of solutions that could not be verified. As a consequence, we also note that \party and \bosy (exp par) now have a higher number of unique solutions. This suggests that these were problems that were solved by \acaciaforaiger and one of the bounded synthesis tools in the realizability track, but in the synthesis track only the solution of the respective bounded synthesis tool could be verified.

Among the tools based on bounded synthesis, \bosy seems to benefit both from the QBF-based encoding ($10$ additional solutions when comparing \bosy (lin) against \party), and from the exponential search strategy ($7$, respectively $10$, additional solutions when comparing the sequential, respectively parallel, versions with linear and exponential search strategy). 

\begin{figure}[h]
\centering
\includegraphics[width=\linewidth]{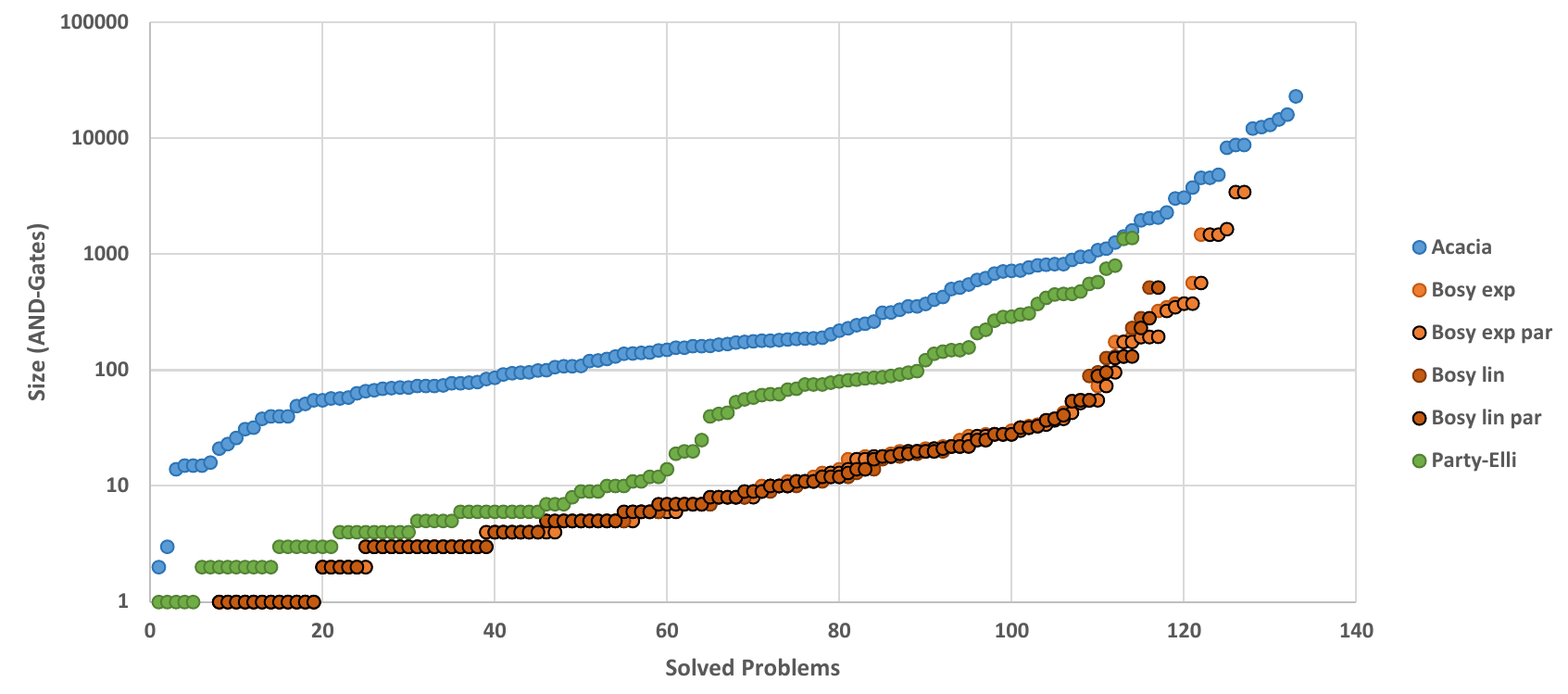}
\caption{TLSF/LTL Synthesis Track: Solution Sizes, based on AND-Gates}
\label{fig:TLSF-size1}
\end{figure}

\begin{figure}[h]
\centering
\includegraphics[width=\linewidth]{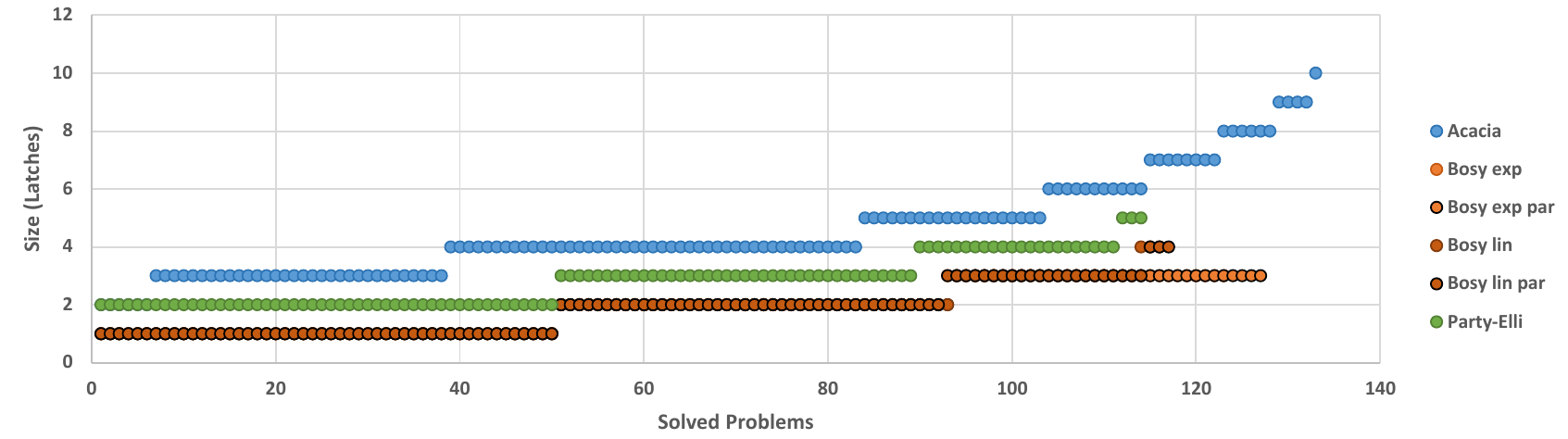}
\caption{TLSF/LTL Synthesis Track: Solution Sizes, based on Latches}
\label{fig:TLSF-size2}
\end{figure}

Even though for \syntcomp 2016 the tools are not ranked with respect to the size of solutions, we want to give a quick comparison. Figure~\ref{fig:TLSF-size1} plots the sizes of solutions based on the number of AND-gates in the AIGER circuit, and Figure~\ref{fig:TLSF-size2} plots the number of latches. Unsurprisingly, we see that solutions of bounded synthesis-based tools are usually smaller than those of \acaciaforaiger. However, there are also some differences in solution size between the solutions of \bosy and \party, pointing out that even if the synthesized solution is minimal with respect to some measure, there may still be differences in the concrete encoding to a circuit. 

%% file: conclusions.tex
\section{Conclusions}
\label{sec:conclusions}

\syntcomp 2016 presented the first major extension of competition setup since the inception of the competition in 2014. In addition to a track based on specifications in the low-level AIGER format for safety properties, the third iteration of \syntcomp included for the first time a track based on specifications in the temporal logic synthesis format (TLSF) for full LTL properties.

In the pre-existing track, two new tools have been entered into the competition, and the results show significant improvements when compared to the best tools of last year. For the new track, three participants entered the competition that was executed for the first time with interesting results regarding the strengths and weaknesses of different existing approaches. 

{ \small
\myparagraph{Acknowledgments}
We thank R\"udiger Ehlers, Ioannis Filippidis, Andrey Kupriyanov, 
Kim Larsen, Nir Piterman, and Markus Rabe for interesting 
suggestions for the future of \syntcomp.

The organization of \syntcomp 2016 was supported by the Austrian Science Fund
(FWF) through project RiSE (S11406-N23) and by the German
Research Foundation (DFG) through project ``Automatic Synthesis of 
Distributed and
Parameterized Systems'' (JA 2357/2-1), and its setup and execution by the 
European Research Council (ERC) Grant OSARES (No.~683300). 

The development of \abssynthe and \acaciaforaiger was supported by an F.R.S.-FNRS fellowship, and
the ERC inVEST (279499) project.

The development of \demiurge was supported by the FWF through project RiSE
(S11406-N23, S11408-N23).

The development of \safetysynth was supported by the 
ERC Grant OSARES (No.~683300). 

The development of \simpleBDD was supported by a gift from the Intel
Corporation.

NICTA is funded by the Australian Government through the
Department of Communications and the Australian Research Council through the ICT
Centre of Excellence Program.
}